# Negative Differential Resistance in Graphene Boron Nitride Heterostructure Controlled by Twist and Phonon-Scattering

Y. Zhao[1], Z. Wan[1], U. Hetmaniuk[2], M. P. Anantram[1]

[1]Department of Electrical Engineering, University of Washington, Seattle, WA 98195, USA

[2]Department of Applied Mathematics, University of Washington, Seattle, WA 98195, USA

Two-dimensional (2D) crystals, such as graphene, hexagonal boron nitride and transitional metal dichalcogenides, have attracted tremendous amount of attention over the past decade due to their extraordinary thermal, electrical and optical properties, making them promising nano-materials for the next-generation electronic systems. A large number of heterostructures have been fabricated by stacking of various 2D materials to achieve different functionalities. In this work, we simulate the electron transport properties of a three-terminal multilayer heterostructure made from graphene nanoribbons vertically sandwiching a boron nitride tunneling barrier. To investigate the effects of the unavoidable misalignment in experiments, we introduce a tunable angular misorientation between 2D layers to the modeled system. Current-Voltage (I-V) characteristics of the device exhibit multiple NDR peaks originating from distinct mechanisms. A unique NDR mechanism arising from the lattice mismatch is captured and it depends on both the twisting angle and voltage bias. Analytical expressions for the positions of the resonant peaks observed in I-V characteristic are developed. To capture the slight degradation of PVR ratios observed in experiments when temperature increases from 2K to 300K, electron-photon scattering decoherence has been added to the simulation, indicating a good agreement with experiment works as well as a robust preservation of resonant tunneling feature.

## I. INTRODUCTION

Unprecedented attention has been brought in two-dimensional (2D) material over the past decades, leading to a variety of van de Waals heterostructures functionalized in both electrical[1-7] and optical[8-10] applications. Prototypical field-effect-transistor (FET) heterostructure devices based on graphene stacked with hexagonal boron nitride (hBN)[11,12] or transitional metal dichalcogenides[4] have been recently realized experimentally. Among these heterostructures, a tunnel-FET device built with hBN vertically sandwiched by two graphene electrodes is of particular interest due to the observation of negative differential resistance (NDR). The appearance of NDR features in this multilayer tunnel-FET structure has the advantages of not requiring a bandgap opening in graphene and relatively simple fabrication process.

Recently, multiple theoretical works have focused on rationalizing the underlying physics of the NDR in graphene-hBN-graphene heterostructures[13-19]. Specifically, two distinct physical mechanisms are responsible for the NDR phenomenon[20], namely the Fabry-Pérot like quantum interference and the bias controlled Dirac cone alignment. These studies assume a perfect "AB" lattice structure between the hBN and graphene sheets. However, the lattice misorientation between stacked 2D atomic crystals is unavoidable during fabrication[12]. In this work, by introducing a tunable angular misorientation between graphene and hBN layers, we investigate the transport properties for a twisted graphene-hBN-graphene device. The current-voltage characteristics exhibit unique NDR features whose essential properties are controlled by the twisting angle and an external gate voltage. By visualizing the misalignment between two Dirac cones in response to the lattice misorientation and bias voltages, we are able to analytically predict the positions of the NDR peaks. To further capture the thermal effects in experiments, a decoherence mechanism, electron-phonon scattering, is introduced and its impact on the NDR effect is presented in good agreement with experiments.

## II. METHODS

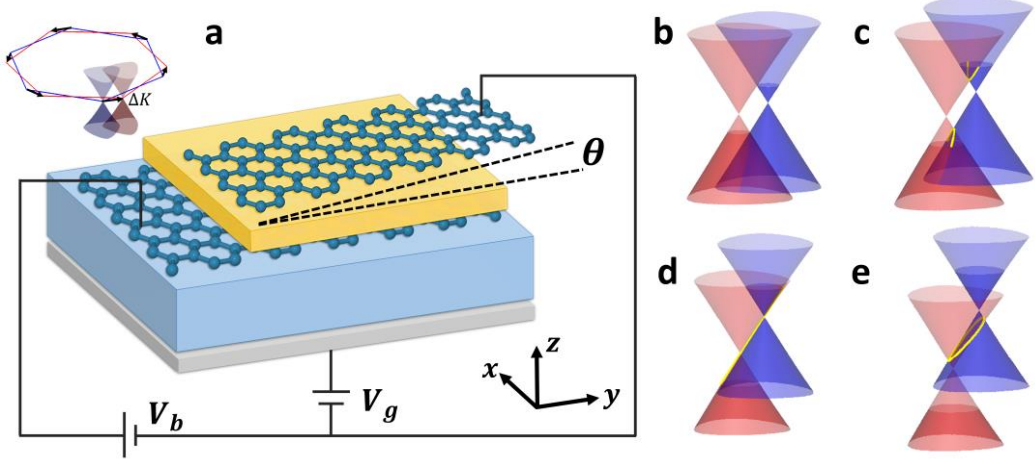

FIG. 1. (a) A schematic view of the twisted heterostructure device. Armchair edged graphene nanoribbon is employed. An external gate electrode is applied on bottom graphene sheet. The top graphene layer is rotated with hBN insulator by an exaggerated angle θ. Inset: the Brillouin zones for bottom and top graphene layers in momentum space. The neutrality points belonging to the same K-valley are displaced by wavevectors whose magnitudes ΔK are identical. (b) – (e) The horizontal distance between neutrality points is determined by the rotation angle θ and the vertical distance between them are determined by the applied gate voltage. Figure (b) depicts the situation of $V_b = V_b^R$. Figures (c) – (e) correspond to situations of $V_b < V_b^P$, $V_b = V_b^P$ and $V_b > V_b^P$. The red and blue cones represent the energy dispersions of bottom and top graphene layers respectively. Occupied and unoccupied states are distinguished by different transparency. The transmissive states that can carry tunnel current is highlighted by yellow curves.

The device (see Figure 1a) consists of two semi-infinitely long monolayer armchair-edged graphene nanoribbon (AGNR) electrodes sandwiching a single layer hBN film as a tunneling barrier[11,12,24]. An external gate electric field is applied vertically to the heterostructure. The multilayer system is stacked in AB order (Bernal stacking) and the lattice constant mismatch between hBN and graphene is ignored. The top graphene layer is rotated by a small tunable angle θ with respect to the central hBN. The sizes of the bottom AGNR and hBN sheets are 22.6nm ($L_x$ along transverse direction) × 13.4nm ($L_y$ along transport direction), and the size of the top AGNR sheet is 14.9nm × 13.4nm. The total number of atoms involved in calculation is 34,109. The system Hamiltonian is constructed by considering a single $p_z$ orbital for C, B and N atoms[21]. We adopt a Slater-Koster model[22] to capture the modulation of the interlayer hopping amplitude due to the lattice misorientation.

The electrostatic model is obtained by considering the device as a three-plate capacitor[12,23], where the quantum capacitance is taken into account for the graphene-plates. Given the values of the bias voltage ($V_b$) and the gate voltage ($V_g$), the chemical potentials of top ($\mu_T$) and bottom ($\mu_B$) AGNR electrodes are determined by solving the following equations:

$$\Delta\varphi_b + \mu_T - \mu_B + eV_b = 0 \tag{1}$$

$$\Delta\varphi_g - \mu_B = eV_g \tag{2}$$

In the first equation, the first term $\Delta\varphi_b = \frac{e^2 d_{BN} n_T}{\epsilon_{BN}}$ is the electrostatic energy difference between graphene electrodes (or equivalently the energy difference between two Dirac points). $d_{BN}$ and $\epsilon_{BN}$ are the thickness and dielectric constant of hBN barrier. $n_{T(B)}$ is the electron concentration in top (bottom) graphene sheet. The second and third terms $\mu_T$ and $\mu_B$ are the chemical potentials of graphene electrodes defined by $\mu_{T(B)} = \pm\hbar v_F \sqrt{\pi|n_{T(B)}|}$ with $v_F$ being the Fermi velocity of graphene. Note that the chemical potential in this paper is defined as the energy difference from the Fermi-level to the Dirac points. In the second equation, $\Delta\varphi_g = \frac{e^2 d_{OX} n_{ext}}{\epsilon_{OX}}$ is the electrostatic energy difference between bottom graphene and gate electrode. $d_{OX}$ is the thickness of gate oxide. $n_{ext}$ denotes the gate-induced charge density on gate electrode (typically *n*-Si), satisfying $n_B + n_T + n_{ext} = 0$. In realistic modeling, the electric field through the gate oxide is mainly determined by the gate voltage, that is $\mu_B \ll eV_g$, reducing equation (2) to $\Delta\varphi_g \sim eV_g$. Therefore, the electrostatic model is governed by the equation:

$$\frac{e^2 d_{BN} n_T}{\epsilon_{BN}} + \mu(n_T) + \mu\left(n_T + \frac{\epsilon_{OX} V_g}{d_{OX}}\right) + eV_b = 0 \tag{3}$$

We have numerically verified that approximating equation (2) only induces a $V_g$ difference smaller than 0.5V, which is ignorable when compared to the typical values of applied $V_g$. This reduced electrostatic model is consistent to references[12,24].

The quantum transport is simulated by using the *non-equilibrium Green's function* (NEGF) method[25]. Instead of solving Green's function by the widely used recursive approach, a novel method, namely *HSC-extension*[26], enables us to efficiently perform the requisite large-scale calculations.

## III. RESULTS

### A. Features of NDR peaks

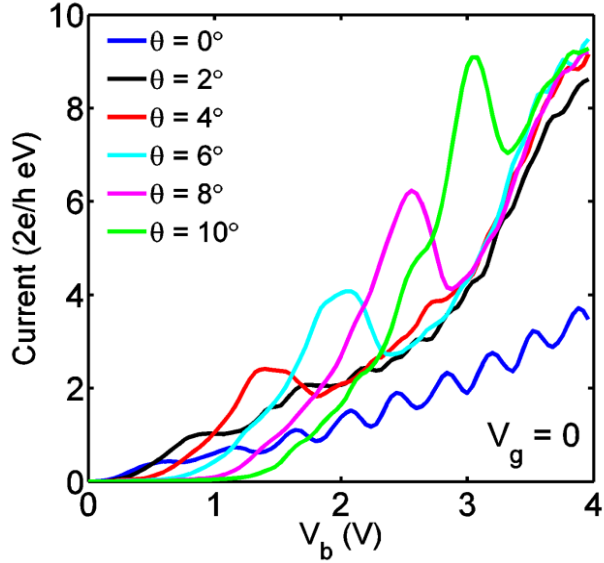

*FIG. 2. Calculated I-V curves for a family of twisting angle without external gate electrode ($V_g = 0$).*

We start by analyzing the twisted device with zero external gate voltage. Figure 2 plots the current-voltage characteristics of the twisted heterostructure with various misorientation angle ($\theta$) of top graphene layer with respect to the hBN layer. The simulated I-V curves exhibit strong NDR peaks, whose location and peak current depends on the misalignment angles.

In the untwisted system ($\theta = 0$), the I-V curves in Figure 2 show multiple current peaks which are fully induced by a Fabry-Pérot interference mechanism[20]. When $\theta$ deviates from perfect alignment and increases, the oscillations gradually disappear. We explain this quenching of current peaks at finite twisting angles by looking at the resonant condition of Fabry-Pérot like interference. In the case of perfect lattice alignment, the transmission states lie on a circular curve with wavevectors at the same energy. When the energy of these states satisfies the resonant condition for Fabry-Pérot interference, all states along the circular curve are capable of carrying current. However, the angular misorientation between graphene layers creates a displacement between two Dirac cones in momentum space. As a result of the conic intersection (Figure 1), the transmission states lie on a hyperbolic or elliptic curve without sharing the same energy. Therefore, the number of states that can tunnel resonantly with the assistance of Fabry-Pérot like interference as well as satisfy the conservation rules are greatly suppressed, thus leading to the current oscillations.

At non-zero $\theta$, the current is close to zero at small biases in Figure 2, and the current rapidly grows after a particular bias voltage $V_b^R$. We explain this feature by depicting the conic dispersions of the two graphene layers at $V_b = V_b^R$ in Figure 1b. When $V_b < V_b^R$, although the Dirac cones intersect along a hyperbolic curve, all transmissive states are occupied in both top and bottom graphene layers. As a result, the tunneling current is close to zero and the device has a high resistance. At the bias voltage $V_b^R$, the occupied / unoccupied states of the bottom (red) / top (blue) Dirac cones intersect only at two points shown in Figure 1b (also in the bottom right inset of Figure 3). When $V_b > V_b^R$, a fraction of the states in the hyperbolic intersection is unoccupied at top layer (Figure 1c), resulting in a rapid increase of current.

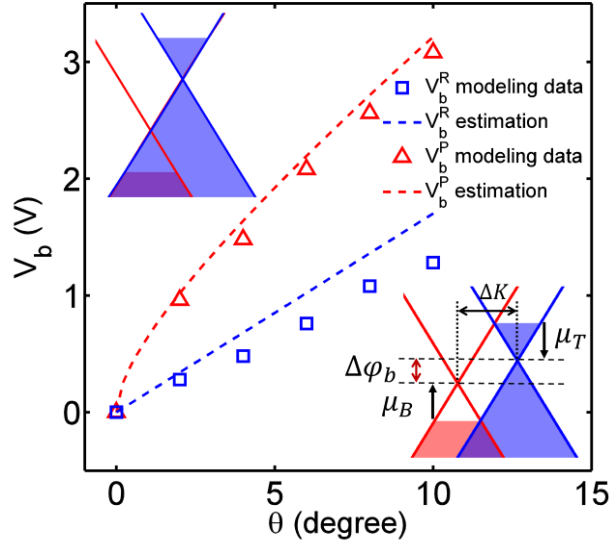

FIG. 3. $V_b^R$ and $V_b^P$ as a function of $\theta$ for both simulated results (Figure 2) and theoretical estimations (eq. 4 and 5). Top inset: illustration of the situation at $V_b^P$, a 2D version of Figure 1d. Bottom inset: illustration of the situation at $V_b^R$ where tunneling current begins to increase rapidly from zero, a 2D version of Figure 1b. In both insets, red and blue cones correspond to energy dispersions for bottom and top graphene layers. The momentum displacement $\Delta K$ between two cones is created by rotation. The meaning of $\Delta \varphi_b$, $\mu_B$ and $\mu_T$ is defined in equation (1).

To explain how $V_b^R$ changes as a function of $\theta$, we provide a formula for $V_b^R$ by solving equation (3) under the situation displayed in Figure 1b (a corresponding 2D illustration can be found in Figure 3 inset), that is $\mu_B + (-\Delta\varphi_b + \mu_B) = \hbar v_F \Delta K$, where $\Delta K = \frac{4\pi}{3a}\theta$. This yields an analytical expression for $V_b^R$ at small $\theta$ ($V_g = 0$):

$$V_b^R = \hbar v_F \Delta K = \frac{4\pi}{3a}\hbar v_F \theta \qquad (4)$$

Next, as the source-drain bias becomes larger, the NDR peak induced occurs at $V_b^P$, which also depends sensitively on twisting angle. Solving equation (3) corresponding to Figure 1d (see 2D illustration in Figure 3 inset), and using $\Delta\varphi_b = \hbar v_F \Delta K$, the value of $V_b^P$ can be expressed as a function of $\theta$:

$$V_b^P = \frac{\hbar v_F}{e}\Delta K + \frac{(\hbar v_F)^{\frac{3}{2}}}{e^2}\sqrt{\frac{\pi\epsilon_0\epsilon_{BN}}{d_{BN}}}\left(\sqrt{\Delta K} + \sqrt{\Delta K - \frac{eV_g d_{BN}}{\hbar v_F d_{OX}}}\right) \qquad (5)$$

The analytical estimations of $V_b^R$ and $V_b^P$ for various $\theta$, as well as the corresponding values extracted from our simulation results, are plotted in Figure 3, which indicates a quantitative match between the analytical formula and numerical results. Note that our calculations consider small misalignment angles (up to $\theta = 10°$), and a larger $\theta$ will result in peak positions deviating from theoretical predictions. Further rotation (for instance, at $\theta = 60°$, $120°$, etc.) will give rise to perfect alignment again, where NDR peaks emerge from the Fabry-Pérot mechanism.

## B. Gate controllability

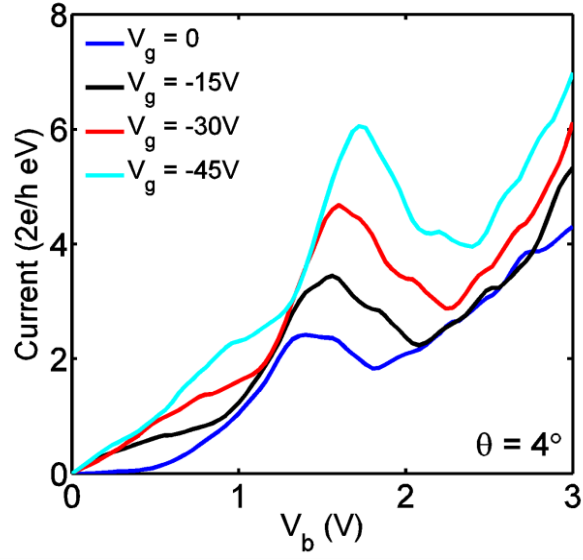

*FIG. 4. Calculated I-V curves for a fixed non-zero twisting angle θ = 4° with various gate voltages.*

For a twisted heterostructure with fixed angle $\theta = 4°$, we model the current-voltage characteristics with various values of $V_g$ in Figure 4. Pronounced resonant peaks whose locations and amplitudes vary as a function of gate voltage are seen as the gate electrode modulates the electrostatic potentials by changing the carrier concentration in graphene layers. As a result, the gate electrode alters the energy difference between the Dirac points on the two sheets, thereby shifting the value of $V_b^P$. Our calculated dependence on $V_g$ is qualitatively consistent with experimental results in reference [12].

## C. Impact of phonon scattering

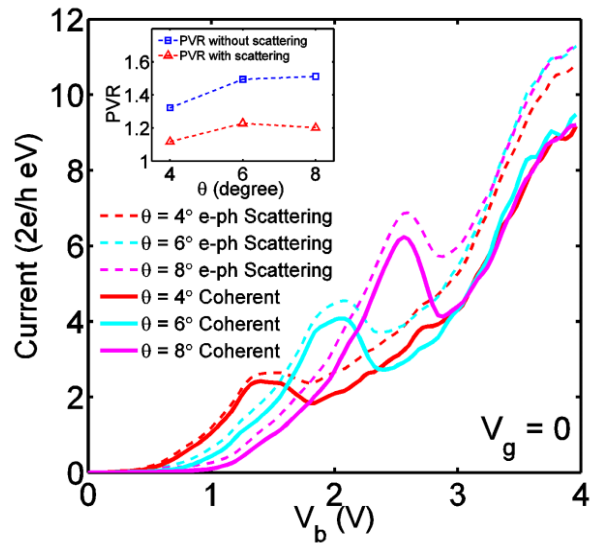

*FIG. 5. Calculated I-V curves for a family of θ at $V_g = 0$ with electron-phonon scattering (dashed curves). For comparison, the corresponding results of coherent transport (from Figure 2) are plotted as solid curves.*

According to ref. [12] when the environmental temperature increases from 2K to room temperature, the measured peak-to-valley ratio (PVR) values are reduced by 10% - 15%. Increasing temperature would increase thermal smearing as well as introduce relaxation / decoherence mechanisms such as electron-phonon scattering. We have verified (results not shown) that when decoherence is *not* present, the difference between I-V curves at low and high temperatures is negligible, indicating that thermal smearing is unlikely to be the dominant mechanism responsible for PVR reduction observed in experiments.

To better interpret the experimental measurements, we have added the electron-phonon scattering in top and bottom graphene nanoribbons within the self-consistent Born approximation[27]. The parameters relevant to the decoherence calculation[28] are: elastic deformation potential $D_{el} = 0.01 \text{eV}^2$, inelastic deformation potential $D_{inel} = 0.07 \text{eV}^2$ and phonon energy $\hbar\omega = 180 \text{meV}$. Phenomenologically, a larger deformational potential reflects stronger electron-phonon scattering and a shorter electron mean free path. The mean free path obtained from our calculations is about $1.42 \mu m$, which is consistent with the measured mean free path of graphene deposited on hBN substrate[29] (around $1.5 \mu m$).

The simulation results with electron-phonon scattering are plotted in Figure 5 (dashed lines). For the twisted heterostructures, the phonon-mediated current as a function of drain voltage preserves the NDR features. The PVR of current magnitude decreases compared to the case of coherent tunneling, whereas the magnitude of both peak and valley current is larger. When electron-phonon scattering exists, the conservation of wavevectors required for the tunneling of electrons between the two layers is weakened, resulting in the rise of tunneling current.

The reduction of PVR values observed in experiment is clearly captured in our simulation. In Figure 5 inset, we plot the PVR values of the current peaks as a function of rotation angle in the coherent case and with phonon-scattering, where a 15% - 25% reduction of PVR values is observed. Therefore, the modeled results are in a reasonable agreement with the observations in experiments, demonstrating that the suppression of NDR features introduced by higher temperature is mainly due to a stronger decoherence mechanism including electron-phonon scattering. Our theoretical work also suggests that for this type of graphene heterostructure devices, keeping the operating temperature low in experiments is critical for observing the features arising from quantum transport.

## IV. Conclusions

In summary, we model the electron transport properties of a three-terminal tunnel-FET device built with twisted graphene layers sandwiching hBN barrier. Robust NDR features in current-voltage characteristics are captured by the numerical simulation and distinct mechanisms are responsible for the resonant tunneling in different situations. The Fabry-Pérot like quantum interference vanishes at larger twisting angles. NDR peaks arising in the case of twisted graphene layers are controllable by both gate voltage and twisting angle. Analytical equations for $V_b^R$ and $V_b^P$ are derived by combining the equation of electrostatic potential with the double Dirac cone model. Moreover, the role of phonon induced decoherence is also numerically simulated to capture the effects of temperature increase in experiments. In the case of twisted graphene sheet, the NDR survives electron-phonon scattering but the peak-to-valley ratios are slightly reduced, consistent with experimental works.

**Acknowledgments**

This work is supported by the National Science Foundation under Grant ECCS-1231927.